\documentclass[twocolumn,trackchanges]{aastex62}
\usepackage{amsmath}
\usepackage{graphicx}
\usepackage{amsfonts}
\usepackage{natbib}
\usepackage{color}
\usepackage{epstopdf}
\usepackage{graphicx}
\usepackage{multirow}
\usepackage{longtable}
\usepackage{rotating}

\shorttitle{Plasma Instability and Measurement of the IGMF}
\shortauthors{Yan et al.}

\begin{document}

\title{Impact of Plasma Instability on Constraint of the Intergalactic Magnetic Field}

\author{Dahai Yan}
\affil{Key Laboratory for the Structure and Evolution of Celestial Objects, Yunnan Observatory, Chinese Academy of Sciences, Kunming 650011, China; yandahai@ynao.ac.cn}
\affil{Center for Astronomical Mega-Science, Chinese Academy of Sciences, 20A Datun Road, Chaoyang District, Beijing 100012, China}
\affil{Department of Astronomy, Key Laboratory of Astroparticle Physics of Yunnan Province, Yunnan University, Kunming 650091, China; zhangpengfee@pmo.ac.cn}

\author{Jianeng Zhou}
\affil{Shanghai Astronomical Observatory, Chinese Academy of Sciences, 80 Nandan Road, Shanghai 200030, China; zjn@shao.ac.cn}

\author{Pengfei Zhang}
\affil{Department of Astronomy, Key Laboratory of Astroparticle Physics of Yunnan Province, Yunnan University, Kunming 650091, China; zhangpengfee@pmo.ac.cn}

\author{Qianqian Zhu}
\affil{Department of General Studies, Nanchang Institute of Science \& Technology, Nanchang 330108, China}

\author{Jiancheng Wang}
\affil{Key Laboratory for the Structure and Evolution of Celestial Objects, Yunnan Observatory, Chinese Academy of Sciences, Kunming 650011, China; yandahai@ynao.ac.cn}
\affil{Center for Astronomical Mega-Science, Chinese Academy of Sciences, 20A Datun Road, Chaoyang District, Beijing 100012, China}



\begin{abstract}

A relativistic electron-positron pair beam can be produced in the interaction of TeV photons from a blazar with the extragalactic background light (EBL).
The relativistic $e^{\pm}$ pairs would loss energy through inverse-Compton scattering (ICS) photons of cosmic microwave background (CMB) or plasma instabilities.
The dominant energy-loss process is under debate.
Based on the assumption that the dominant energy-loss process is ICS, the resulted cascade GeV radiation is usually used to constrain the intergalactic magnetic field (IGMF).
Here, we include the energy-loss due to plasma oblique instability in the calculation of cascade gamma-ray flux, 
and investigate the impact of the plasma instability on the constraint of IGMF.
The up-to-date GeV data and archival TeV data of the blazar 1ES 0229+200 are used.
The results indicate that even if the oblique instability cooling is dominating over ICS cooling, the cascade flux could be still used to constrain the IGMF.
It is found that with the ratio between the cooling rates of the oblique instability and the ICS varying from 0.1, 1 to 10, 
the lower limit of the IGMF putted by the cascade flux and the gamma-ray data changes from $8\times10^{-18}\ $G, $5\times10^{-18}\ $G to $10^{-18}\ $G.
If the ratio between the two cooling rates is 30, the estimate of IGMF based on the cascade flux is invalid.

\end{abstract}



\section{Introduction} \label{sec:intro}

Extragalactic TeV sky is dominated by blazars.
TeV photons from blazars interact with photons of the extragalactic background light (EBL) to create $e^{\pm}$ pairs.
These pairs are relativistic with the Lorentz factors of $\gamma\sim10^6-10^7$.

It is usually assumed that the dominant energy-loss process of the pairs is inverse-Compton scattering (ICS) photons of cosmic microwave background (CMB) 
to produce secondary cascade GeV emissions \citep[e.g.,][]{Aharonian,Dai,Fan}.
The intergalactic magnetic field (IGMF) can deflect these $e^{\pm}$ pairs, 
and then modulate the secondary GeV emissions.
Therefore,  gamma-ray astronomy is thought to provide a useful probe to the IGMF through investigations on cascade-radiation  spectra of blazars \citep[e.g.,][]{Murase,Neronov,Tavecchio,Taylor,Dermer,Finke}, 
cascade-radiation contribution to extragalactic diffuse gamma-ray background (EGRB) \citep[e.g.,][]{Yan, Venters}, and gamma-ray halos \citep[e.g.,][]{Elyiv,Ackermann,Ackermann18,Chen}.

However, it was recently proposed that as the blazar-induced pair beam moves through the intergalactic medium
(IGM), plasma instabilities would be triggered \citep[e.g.,][]{Bret}. 
As a result, the energy-loss of the pairs would be dominated by plasma
instabilities, thereby suppressing ICS \citep[e.g.,][]{Broderick}.
In this case, the kinetic energy of the pairs are efficiently converted into heat in the IGM, rather than GeV radiations.
This would rewrite thermal history of the IGM \citep[e.g.,][]{Chang,Pfrommer}.
\citet{Broderick} argued that the measurements of the IGMF through gamma-ray emissions would be invalidated in this situation.

So far, the fate of the pairs evolution is still under debate \citep[e.g.,][]{Schlickeiser12a,Schlickeiser12b,Schlickeiser13,Miniati,Chang14,Sironi,Kempf,Shalaby,Vafin}.
These authors studied plasma instabilities through analytical methods and particle-in-cell (PIC) simulations.
In this work, we aim to investigate the impact of the plasma instabilities on measurement of IGMF.

This paper is organized as follows: In Section~\ref{data}, we show results of {\it Fermi}-LAT data analysis for 1ES 0229+200. 
In Section~\ref{implication}, we give our main results. 
Conclusions are given in Section~\ref{4}.

\section{Fermi-LAT data analysis}
\label{data}

The TeV blazar 1ES 0229+200, at redshift z = 0.1396, was observed with HESS in 2006 and 2007 \citep{Aharonian07} and VERITAS in 2009 - 2012 \citep{Aliu}.
Hints of moderate variability on yearly and monthly timescales were found in its TeV emissions \citep{Aliu,Cologna}.
The observed TeV spectrum extends to $\sim$10 TeV, with the photon index $\Gamma_{\rm TeV}\sim2.6$.
The EBL-corrected TeV spectrum has the photon index $\Gamma_{\rm TeV}\sim1.5$ \citep[e.g.,][]{Finke}

At GeV energies, weak
emissions from 1ES 0229+200 have been detected by {\it Fermi}-LAT.
We reanalyze the {\it Fermi}-LAT  data for 1ES 0229+200 collected from 2008 August
to 2018 August, with energies between 100 MeV and 300 GeV. 
The events of PASS8 {\tt SOURCE} class, within a $20^{\circ} \times 20^{\circ}$ region of interest (ROI) centered at 1ES 0229+200, are used. 
Analysis is performed with the ScienceTools v10r0p5 and the instrument response function (IRF) {\tt P8R2\_SOURCE\_V6}.
To avoid contamination from the Earth's albedo, we exclude the events with the zenith angles $> 90^{\circ}$. 
Furthermore, we exclude data when the source was within the $15^{\circ}$ degrees region of the Sun and moon, 
since the gamma-ray emission from 1ES 0229+200 is possibly affected by $\gamma$-rays from the Sun and moon\citep[e.g.,][]{Finke,Inoue}.

\begin{figure*}
 \centering
 \includegraphics[scale=0.5]{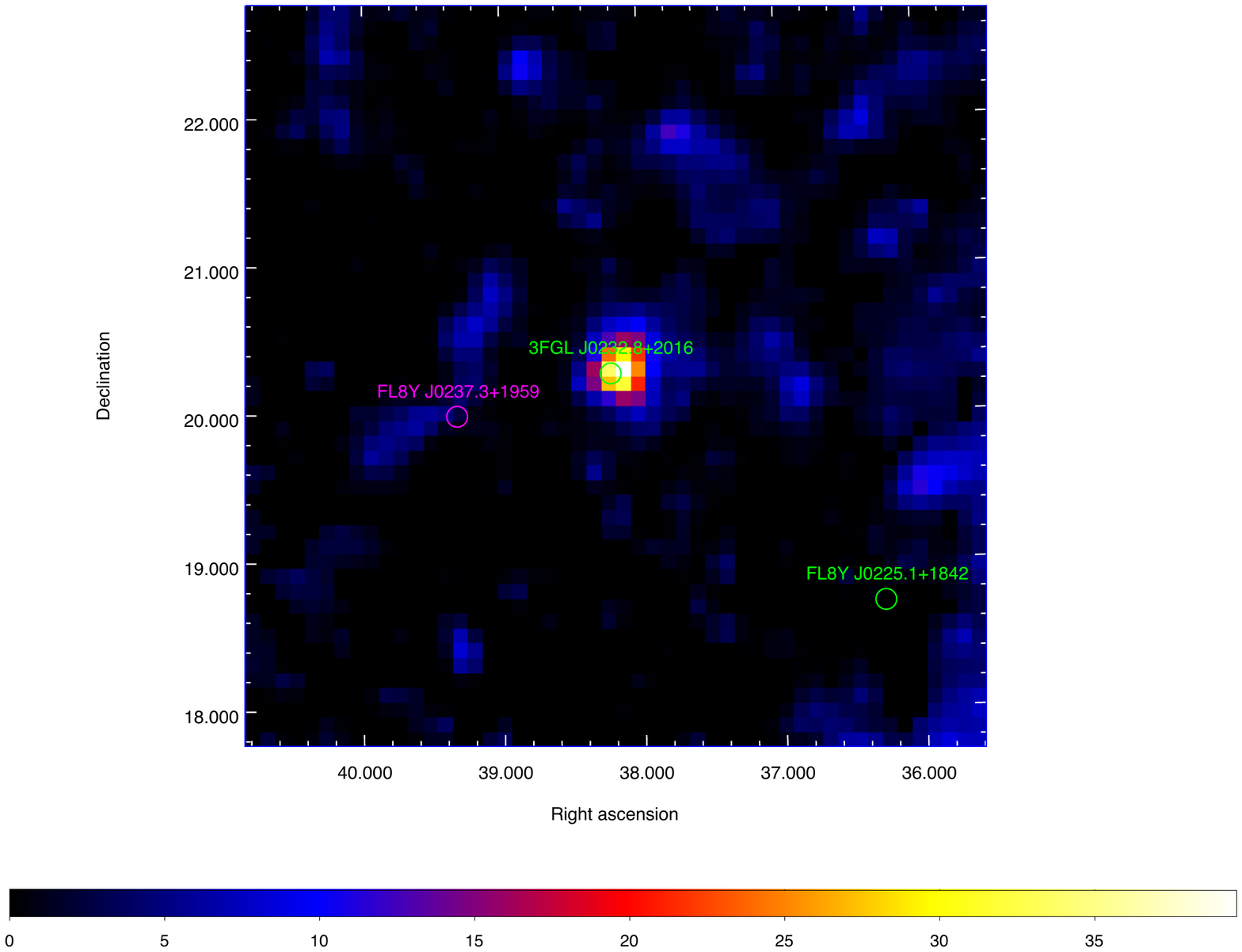}
\caption{$5^{\circ} \times 5^{\circ}$ TS map centered at 1ES 0229+200, with energies between 100 MeV and 10 GeV. 
Two FL8Y sources, FL8Y J0237.3+1959 and FL8Y J0225.1+1842, are shown.}
\label{tsmap}
\end{figure*}

The standard likelihood analysis by {\tt gtlike} is performed through this work. 
In addition to Galactic and extragalactic gamma-ray diffuse background components, the background we consider includes all sources in 3FGL\citep{2015ApJS..218...23A} within the ROI and 
a preliminary LAT 8-year source list\footnote{https://fermi.gsfc.nasa.gov/ssc/data/access/lat/fl8y/} (see Figure~\ref{tsmap}). 
According to the global fitting results, parameters of all sources (except for target source and prefactors of variable sources\footnote{Variable sources are defined as the ones having {\tt Variability\_Index $>$ 72.44} \citep{2015ApJS..218...23A}. Since there is no variability information for FL8Y source yet, and these FL8Y sources in ROI are faint, 
we fix their parameters in calculating light curve. 
Four FL8Y sources involved here are FL8Y J0225.1+1842, FL8Y J0230.2+1714, FL8Y J0237.3+1959 and FL8Y J0242.8+1733. 
\cite{Paliya} analyzed the {\it Fermi}-LAT data of FL8Y J0225.1+1842.}) are fixed to the best-fitting values in constructing light curves. 
For spectral energy distribution analysis, models are just modified by fixing the spectral shapes, i.e., remaining the prefactors free. 
Upper limits at 95\% confidence level are given when the flux with TS $<$ 4 is obtained.

\begin{figure*}
   \centering
     \includegraphics[scale=0.4]{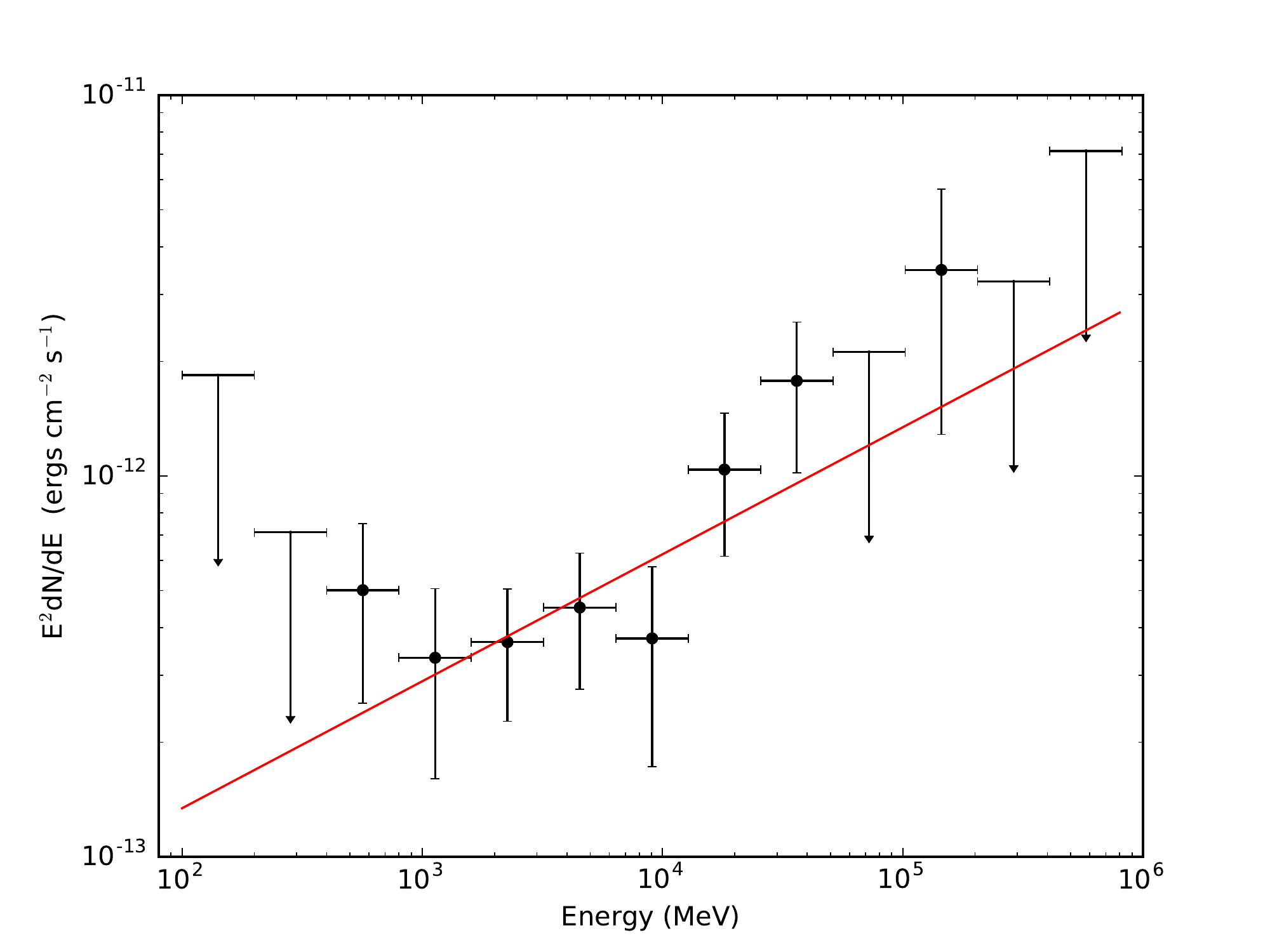}
\caption{Average {\it Fermi}-LAT spectrum for 1ES 0229+200. The solid line is the best-fitting result to the spectrum.} 
\label{sed}
\end{figure*}

\begin{figure*}
 \centering
     \includegraphics[scale=0.4]{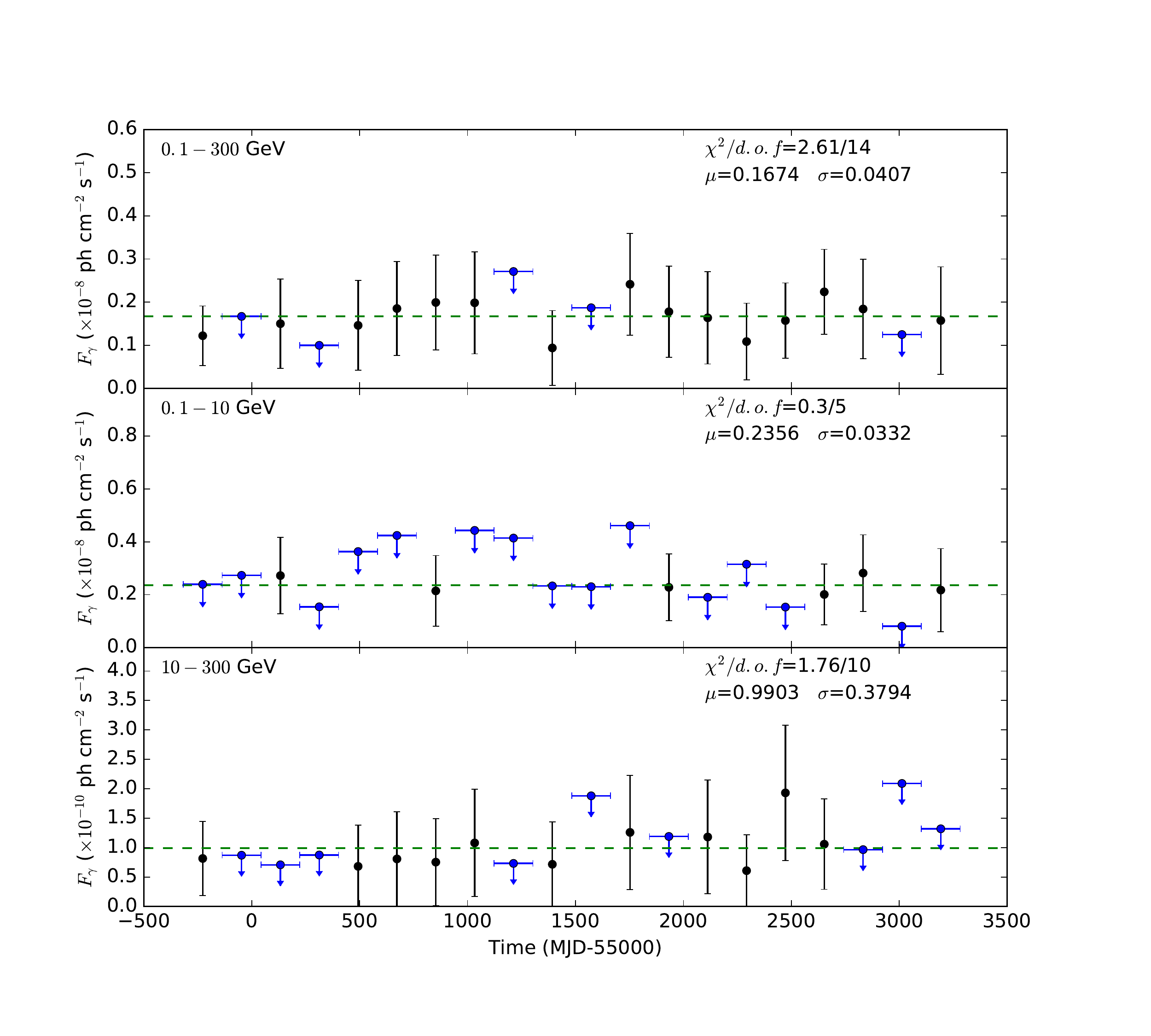}
\caption{ $\gamma$-ray light curves for 1ES 0229+200. Each light curve is fitted by a constant (dashed line), and the best-fitting results are shown in the figure.} 
\label{alllc}
\end{figure*}

In Figure~\ref{sed}, we show the ten-year average {\it Fermi}-LAT spectrum for 1ES 0229+200.
This spectrum can be described by a power-law function, with the photon index $\Gamma_{\rm LAT}=1.67\pm0.11$.
No variability is found in the LAT data (Figure~\ref{alllc}).

\section{Results}
\label{implication}

\subsection{Calculation of cascade gamma-ray flux including plasma instability cooling}

The steady-state electron
continuity equation that governs the pair evolution is\footnote{We reasonably assume no escape and advection, like \citet{Broderick}.}

\begin{equation}
\label{kineticEQ}
\frac{\partial}{\partial{\gamma}}\left[\dot{\gamma} n_e (\gamma)\right]=\dot{n}_e (\gamma)\ , 
\end{equation}
where $\dot{\gamma}$ is the energy-loss of the pairs and $\dot{n}_e (\gamma)$ is the injection rate of the pairs.
The solution to this equation is 
\begin{equation}
\label{solution}
n(\gamma)=\frac{1}{\dot{\gamma}}\int^\infty_{\gamma}d\gamma'\dot{n}_e (\gamma')\ . 
\end{equation}
The pairs injection rate is \citep{Dermer13}
\begin{equation}
\label{inj}
\dot{n}_e (\gamma)=\frac{16\pi d_{L}^2f_{\epsilon}}{m_ec^2\epsilon^2}[1-e^{-\tau_{\gamma\gamma}(\epsilon,z)}],\ \epsilon=2\gamma\ ;
\end{equation}
where $d_L$ is the luminosity distance and $\tau_{\gamma\gamma}(\epsilon,z)$ is the EBL absorption depth for gamma-rays with energy of $\epsilon m_ec^2$ emitted at redshift $z$.
The primary $\nu F_{\nu}$ gamma-ray spectrum is 
\begin{equation}
\label{inj}
f_\epsilon=F(E)=F_0(\frac{E}{E_0})^{2-\Gamma}e^{-E/E_{\rm cut}},\ E=m_ec^2\epsilon\ ;
\end{equation}
where $\Gamma$ is photon index of the primary spectrum, $F_0$ is a normalization flux, and $E_{\rm cut}$ is a high-energy cut-off.
$f_{\epsilon}e^{-\tau_{\gamma\gamma}(\epsilon,z)}$ is the EBL-absorbed primary spectrum.

We consider the energy-losses of the pairs due to ICS and plasma instabilities.
The energy-loss rate due to inverse-Compton scattering photons of CMB in the Thomson regime is
\begin{equation}
\label{ICTH}
-\dot{\gamma}_{\rm T}=\frac{4}{3}c\sigma_{\rm T}\frac{u_{\rm CMB}}{m_ec^2}\gamma^2=\nu_{\rm T}\gamma^2\ ,
\end{equation}
where $u_{\rm CMB}=4\times10^{-13}\rm\ erg\ cm^{-3}$ is the CMB energy density at low redshifts.
For the energy-loss due to plasma instabilities, 
we consider the oblique instability which has the most powerful growth \citep[e.g.,][]{Bret}.
Its energy-loss rate can be simply written as \citep{Broderick}
\begin{equation}
\label{mk}
-\dot{\gamma}_{\rm M,K}=\nu_{\rm M,k}\gamma^2\ .
\end{equation}

The total energy-loss rate is $\dot{\gamma}=\dot{\gamma}_{\rm M,K}+\dot{\gamma}_{\rm T}=\nu_{\rm T}(1+b)\gamma^2$, where $b=\nu_{\rm M,k}/\nu_{\rm T}$.
Using this total energy-loss rate and the formula of \cite{Dermer13}, we calculate the cascade GeV flux $F_{\rm cas}$.

In \cite{Dermer} and \cite{Dermer13} , the calculation of the cascade flux only considered the energy-loss due to ICS.
As we show above, when the energy-loss due to the oblique instability is taken into account, the cascade flux in \cite{Dermer} and \cite{Dermer13} is simply modified by a factor of $1/(1+b)$.

In addition to the parameters of the primary spectrum, the other parameters that can effect cascade flux are IGMF strength $B_{\rm IGMF}$, IGMF coherence length $\lambda_{B}$,
jet opening angle $\theta_{\rm j}$, and the blazar active time with constant flux $t_{\rm live}$.

\subsection{Constraining $B_{\rm IGMF}$ in different cooling regimes}

In the following calculations, we choose $\theta_{\rm j}=$0.1 rad, $\lambda_{B}=$1 Mpc, $t_{\rm live}=10$ years, $E_{\rm cut}=$5 TeV, $\Gamma=1.4$, 
$F_0=3\times10^{-11}\rm\ erg\ cm^{-1}\ s^{-1}$,
and the best-fit EBL model of \cite{Kneiske}.
 
\begin{figure*}
 \centering
     \includegraphics[scale=0.5]{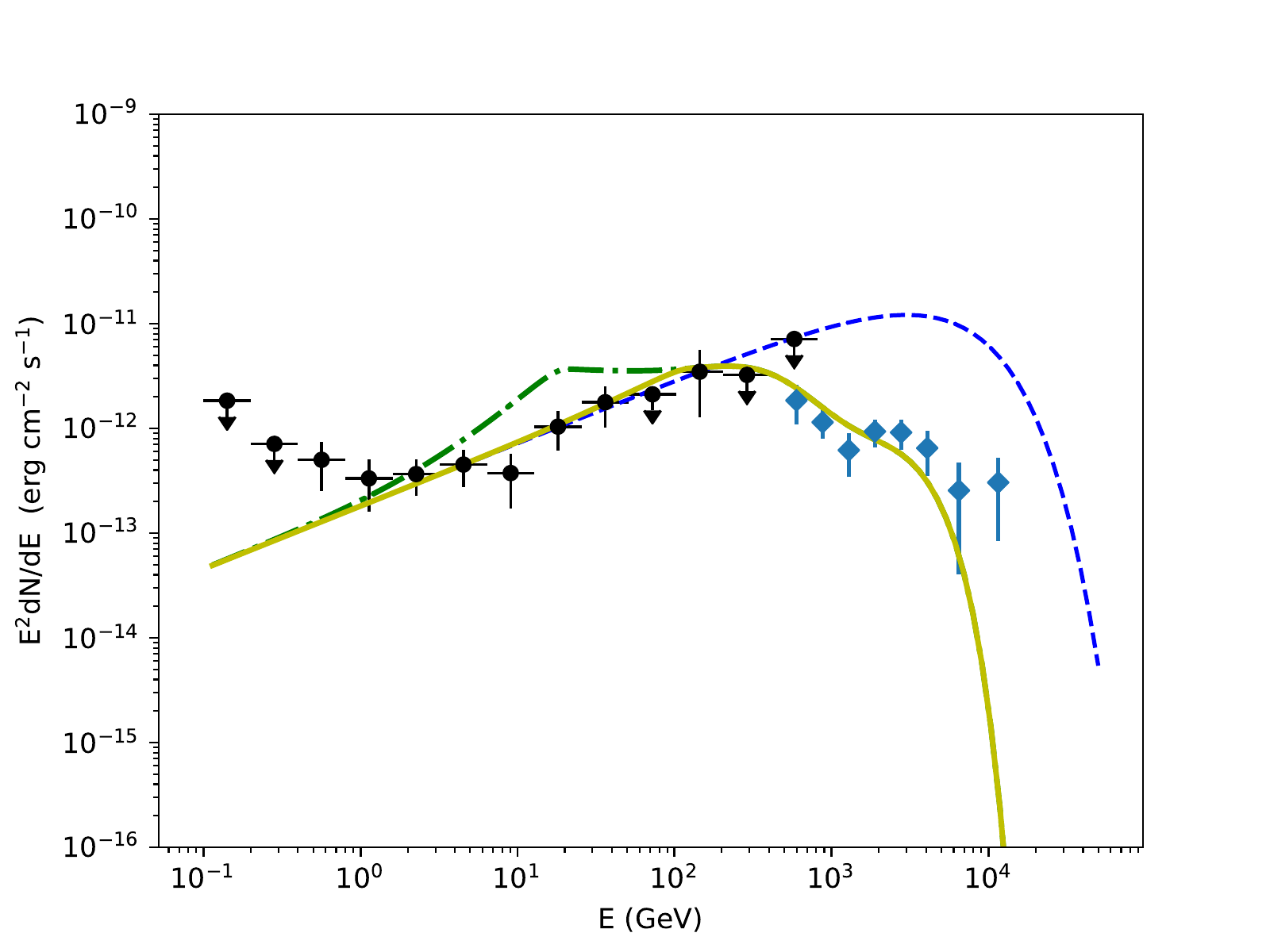}
\caption{Reproducing results of LAT spectrum (circles) and HESS spectrum (diamonds) with the ``primary+cascade'' model,
             in the case of ICS cooling dominating over the oblique instability cooling ($b=0.1$).
             The dashed line is the intrinsic primary gamma-ray spectrum.
             The dash-dotted and solid lines are the sum of cascade flux and primary flux after EBL absorption respectively with $B_{\rm IGMF}=$$10^{-18}\ $G 
             and  $B_{\rm IGMF}=$$8\times10^{-18}\ $G . $t_{\rm live}=10\ $years is used in the calculations.} 
\label{c1}
\end{figure*}

 We first consider the case of ICS cooling dominating over the oblique instability cooling through using $b=0.1$.
 The result is shown in Figure~\ref{c1}.
 One can see that in order to avoid the cascade flux exceeding the LAT flux, $B_{\rm IGMF}$ is required to be larger than $8\times10^{-18}\ $G.
 
\begin{figure*}
 \centering
     \includegraphics[scale=0.5]{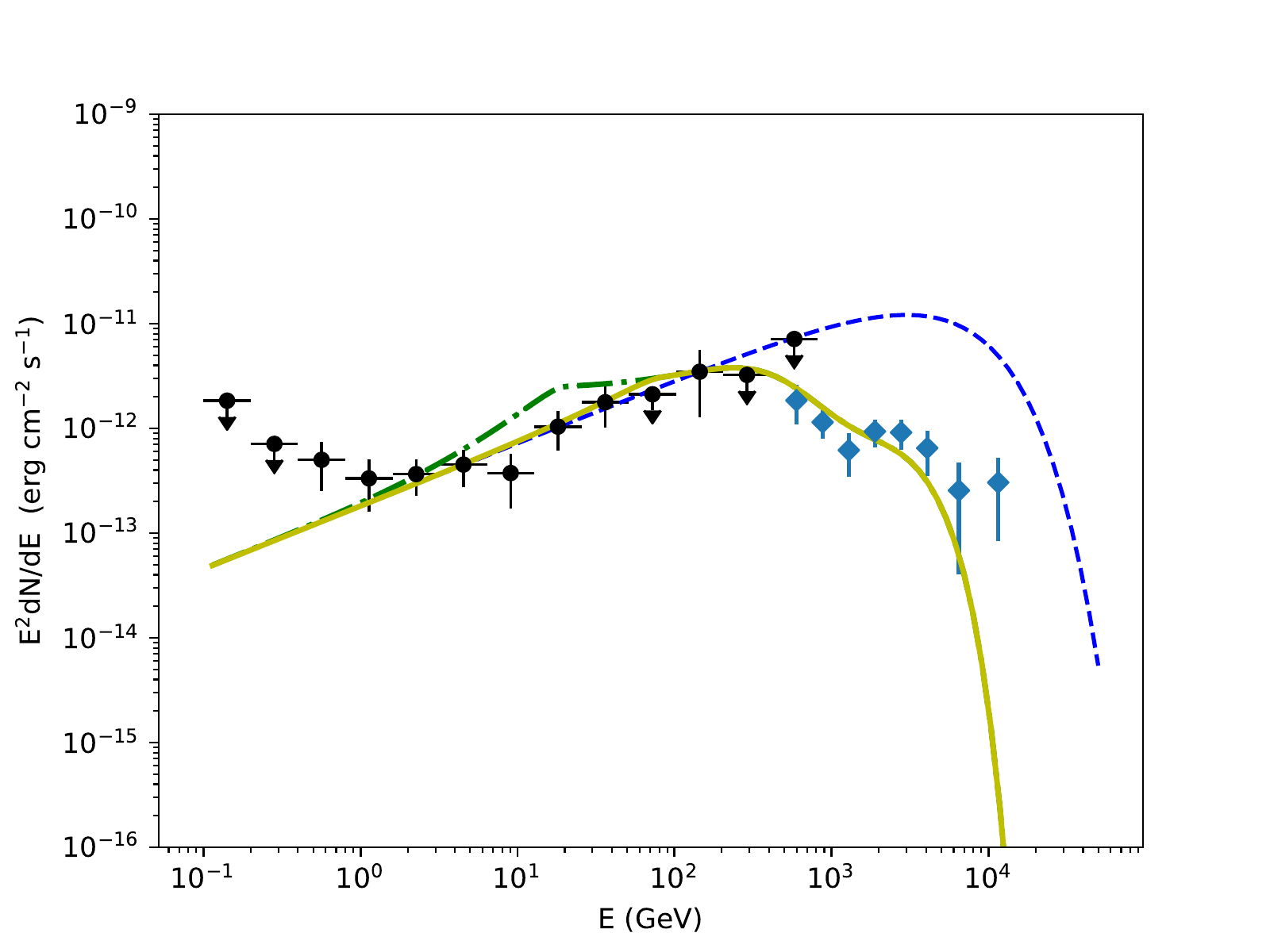}
\caption{Same as Figure~\ref{c1}, but in the case of the oblique instability cooling comparable with  the ICS cooling ($b=1$).
             Dash-dotted line:  $B_{\rm IGMF}=$$10^{-18}\ $G; solid line: $B_{\rm IGMF}=$$5\times10^{-18}\ $G.} 
\label{c2}
\end{figure*}

\begin{figure*}
 \centering
     \includegraphics[scale=0.5]{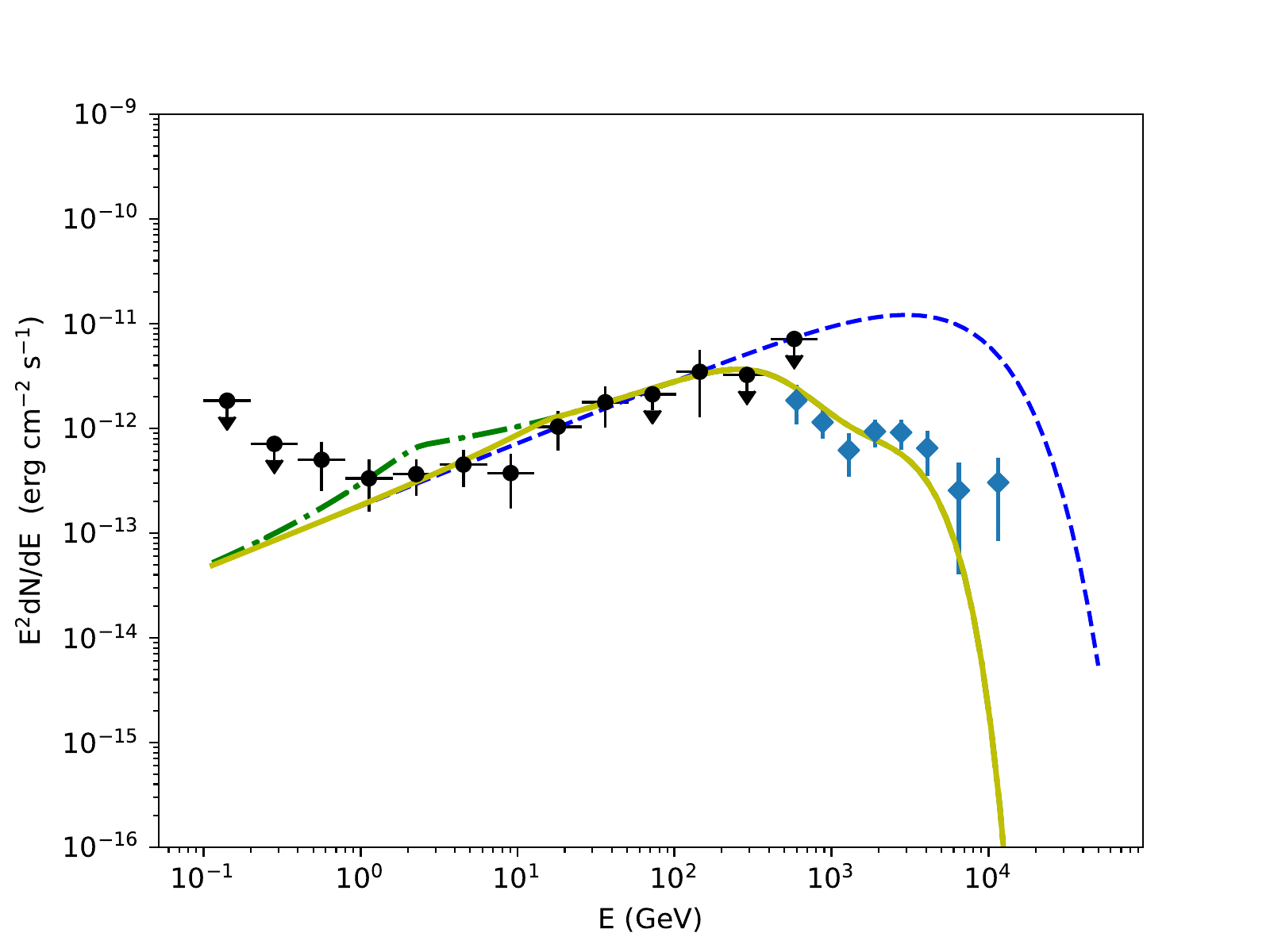}
\caption{Same as Figure~\ref{c1}, but in the case of the oblique instability cooling dominating over  the ICS cooling ($b=10$).
             Dash-dotted line:  $B_{\rm IGMF}=$$10^{-19}\ $G; solid line: $B_{\rm IGMF}=$$10^{-18}\ $G.} 
\label{c3}
\end{figure*}

 With $b=1$ and 10, we obtain $B_{\rm IGMF}\ge5\times10^{-18}\ $G (Figure~\ref{c2}) and $\ge10^{-18}\ $G (Figure~\ref{c3}) , respectively.
 
 In Figure~\ref{c3}, we give the results with $b=30$. 
 It is found that the contribution of the cascade flux to LAT data is negligible even an extremely low $B_{\rm IGMF}$ ($10^{-24}\ $G) is assumed.
 
\begin{figure*}
 \centering
     \includegraphics[scale=0.5]{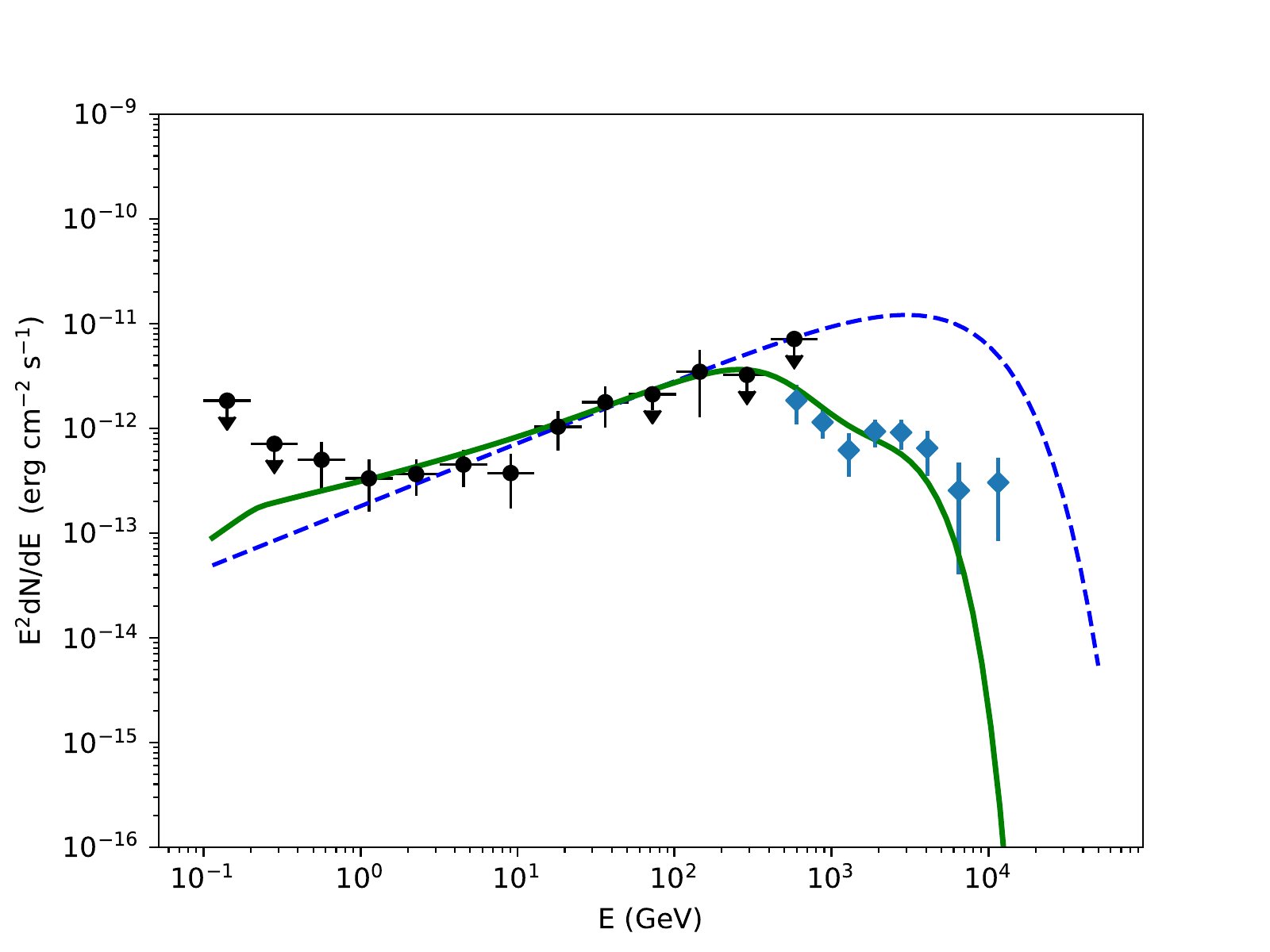}
\caption{Same as Figure~\ref{c1}, but with $b=30$.
             Solid line: $B_{\rm IGMF}=$$10^{-24}\ $G.} 
\label{c4}
\end{figure*}

\section{CONCLUSIONS}
\label{4}

Given that plasma instabilities may play an important role in the evolution of a blazar-induced pair beam \citep[e.g.,][]{Broderick,Shalaby,Vafin}, 
we take the plasma oblique instability (with the fastest growth)  cooling into account in considering the evolution of relativistic electron-positron pairs created in propagation of TeV photons from a blazar.
This would alter the density of pairs per unit Lorentz factor, and then the cascade gamma-ray flux produced by the pairs ICS CMB photons.

Using the up-to-date {\it Fermi}-LAT
observations and archival HESS observations of 1ES 0229+200, 
we constrain IGMF strength $B_{\rm IGMF}$ in different cooling processes.
The results suggest that the gamma-ray data still put effective constraint on $B_{\rm IGMF}$, 
even if the oblique instability cooling strongly dominating over the ICS cooling, 
e.g., the ratio between the cooling rates of the oblique instability and the ICS $b$ $\sim10$.
We find that with this ratio varying from 0.1, 1 to 10, 
the lower limit of $B_{\rm IGMF}$ putted by the cascade flux and gamma-ray data changes from $8\times10^{-18}\ $G, $5\times10^{-18}\ $G to $10^{-18}\ $G.
From the ICS cooling dominating to the oblique instability cooling dominating, the lower limit of $B_{\rm IGMF}$ changes by a factor of $\sim10$.

Compared with that obtained in the case of ICS cooling dominating, the lower limit of $B_{\rm IGMF}$ is slightly changed in the case of the two cooling processes comparable. 

It is also noted that with $b=30$, the cascade flux is negligible compared with the LAT data.
This means the gamma-ray astronomy cannot be considered as a useful probe to IGMF any more.

In Appendix, we perform the analyses with the EBL model (``Model C") of \cite{Finke10}
\footnote{This EBL model is similar to the EBL models of \cite{Franceschini} and \cite{Dom}.}.
The results (Figure~\ref{eblF}) are quite similar to that calculated by using the EBL model of \cite{Kneiske}.

\section*{Acknowledgements}
We thank the referee for the constructive suggestions. 
We acknowledge financial supports from the National
Natural Science Foundation of China (NSFC-11803081, NSFC-11573060, NSFC-11573026, NSFC-U1738124, NSFC-11603059 and
NSFC-11661161010) and the joint foundation of Department of Science and Technology of Yunnan Province and Yunnan University [2018FY001(-003)].
The work of D. H. Yan is also supported by the CAS
``Light of West China'' Program.
\bibliography{0229}

\appendix
\label{app}
\section{Impact of EBL uncertainty on the results}
\begin{figure*}
 \centering
     \includegraphics[scale=0.4]{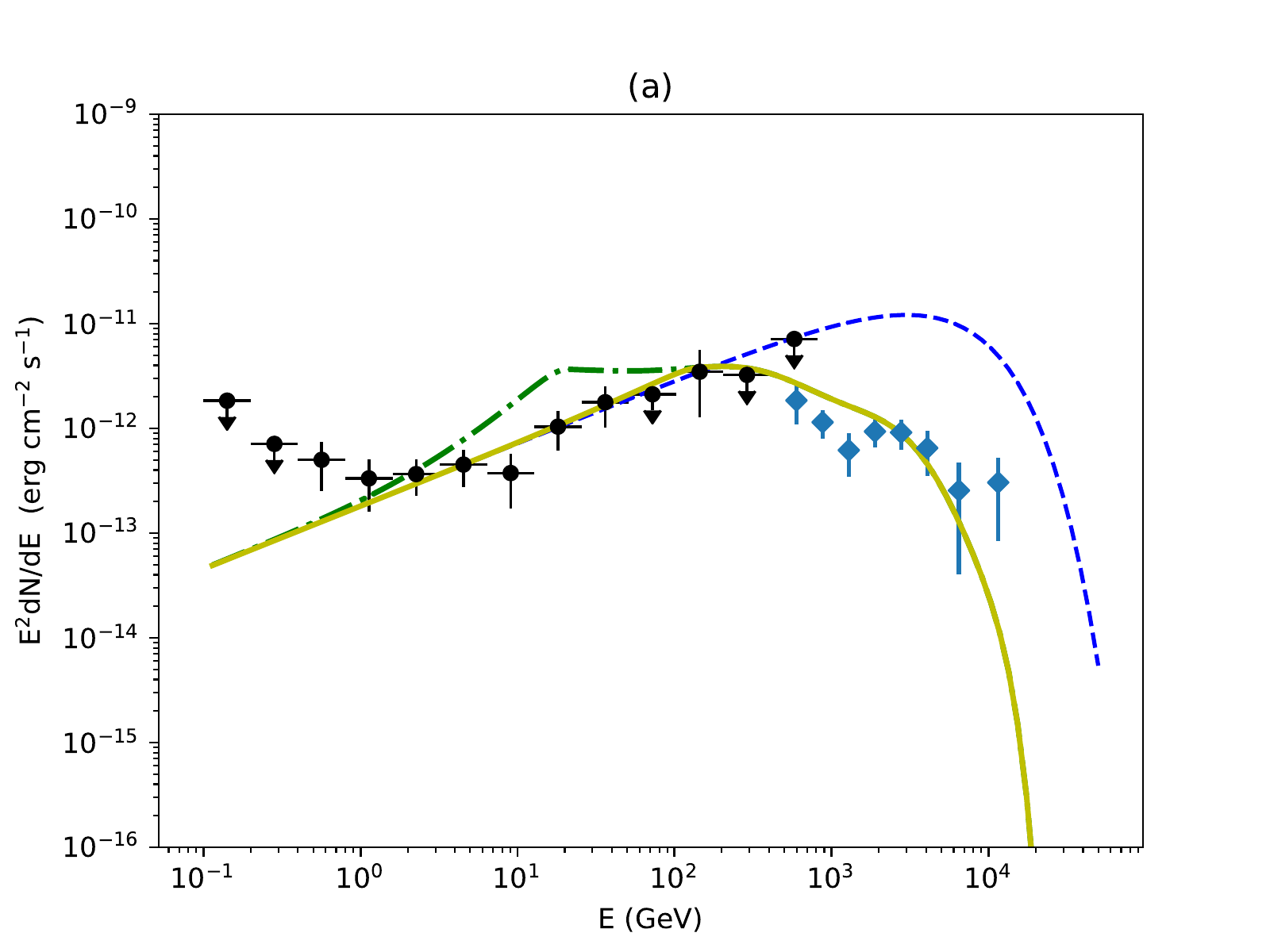}
     \includegraphics[scale=0.4]{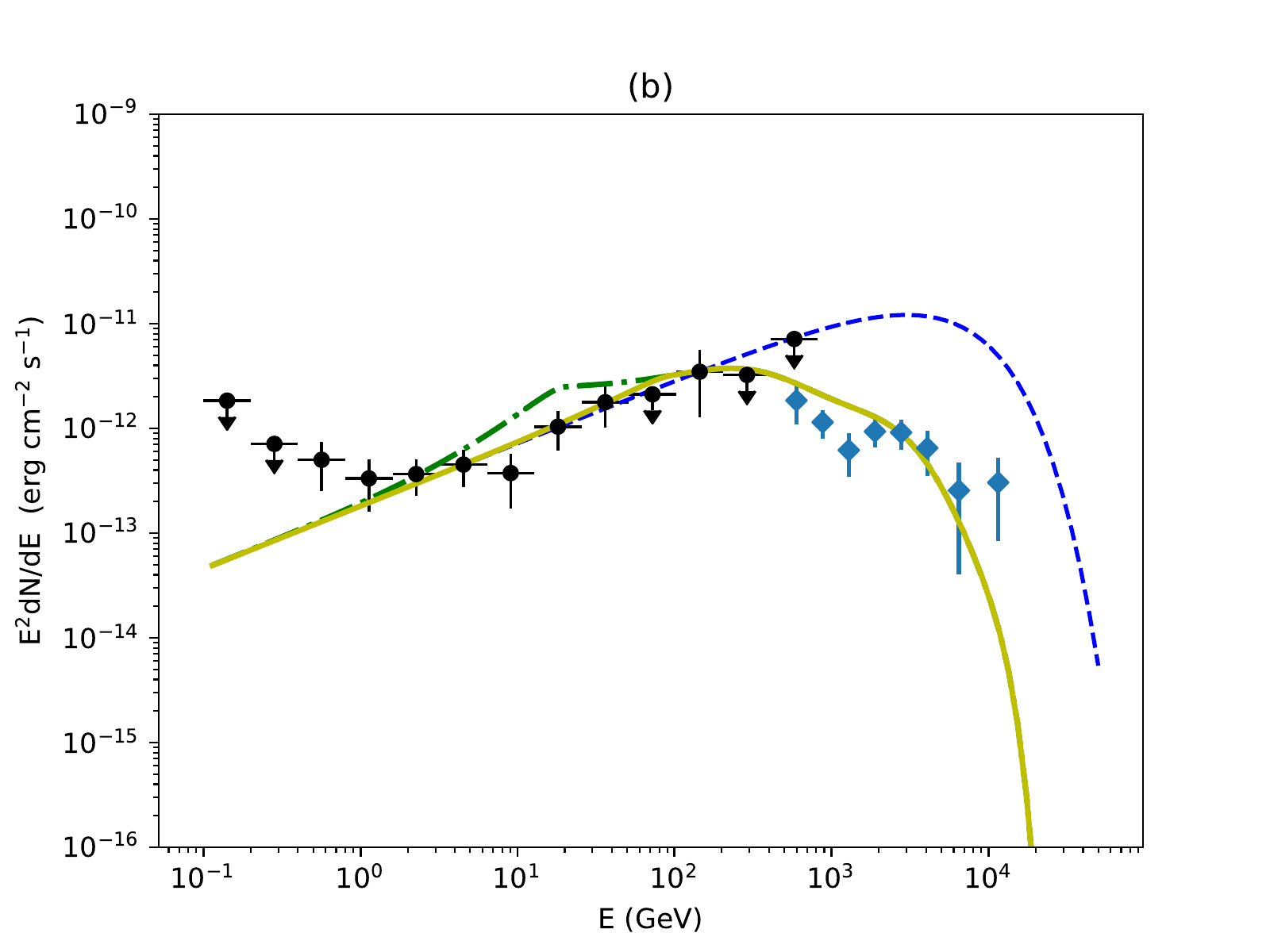}
     \includegraphics[scale=0.4]{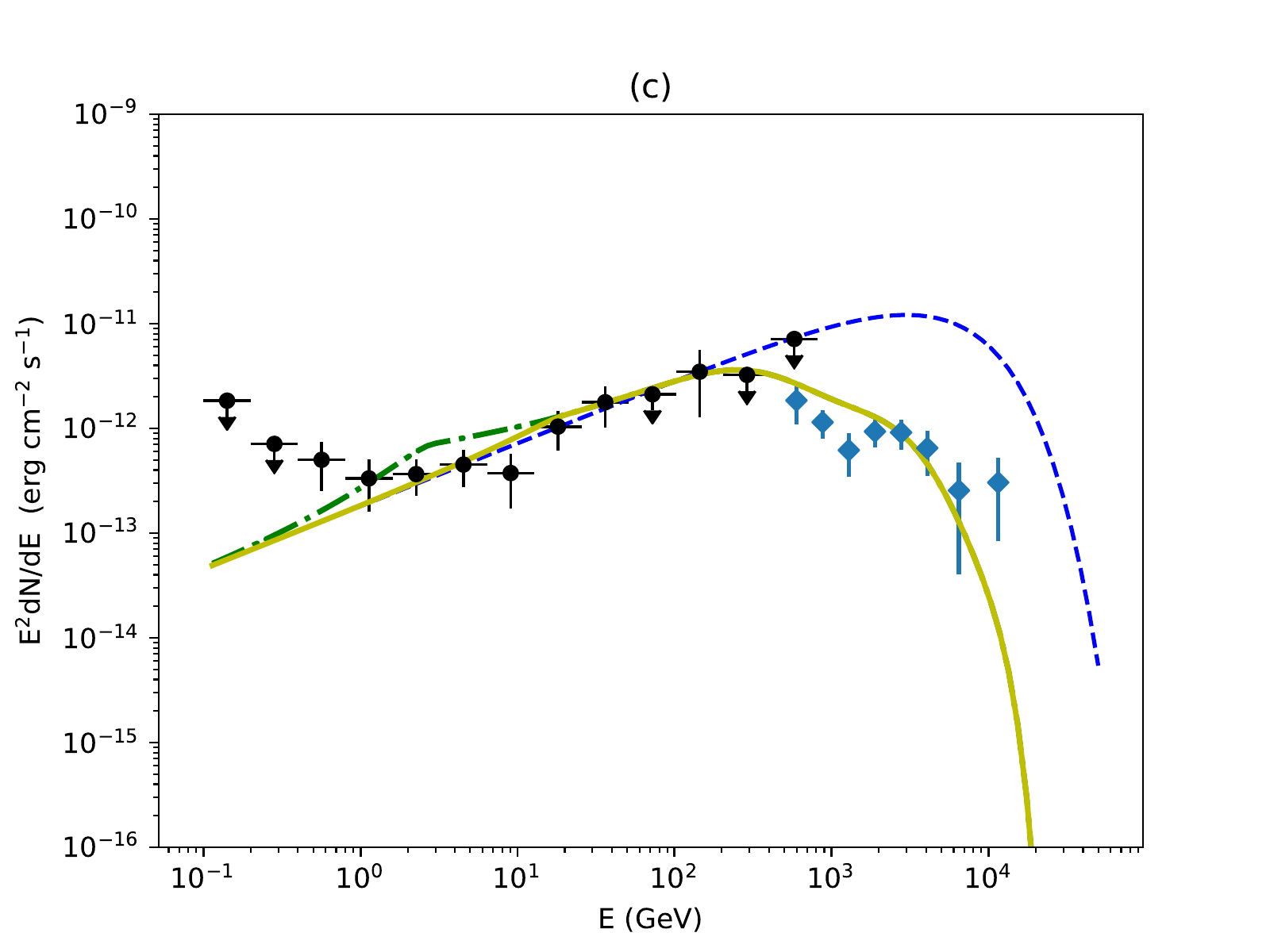}
     \includegraphics[scale=0.4]{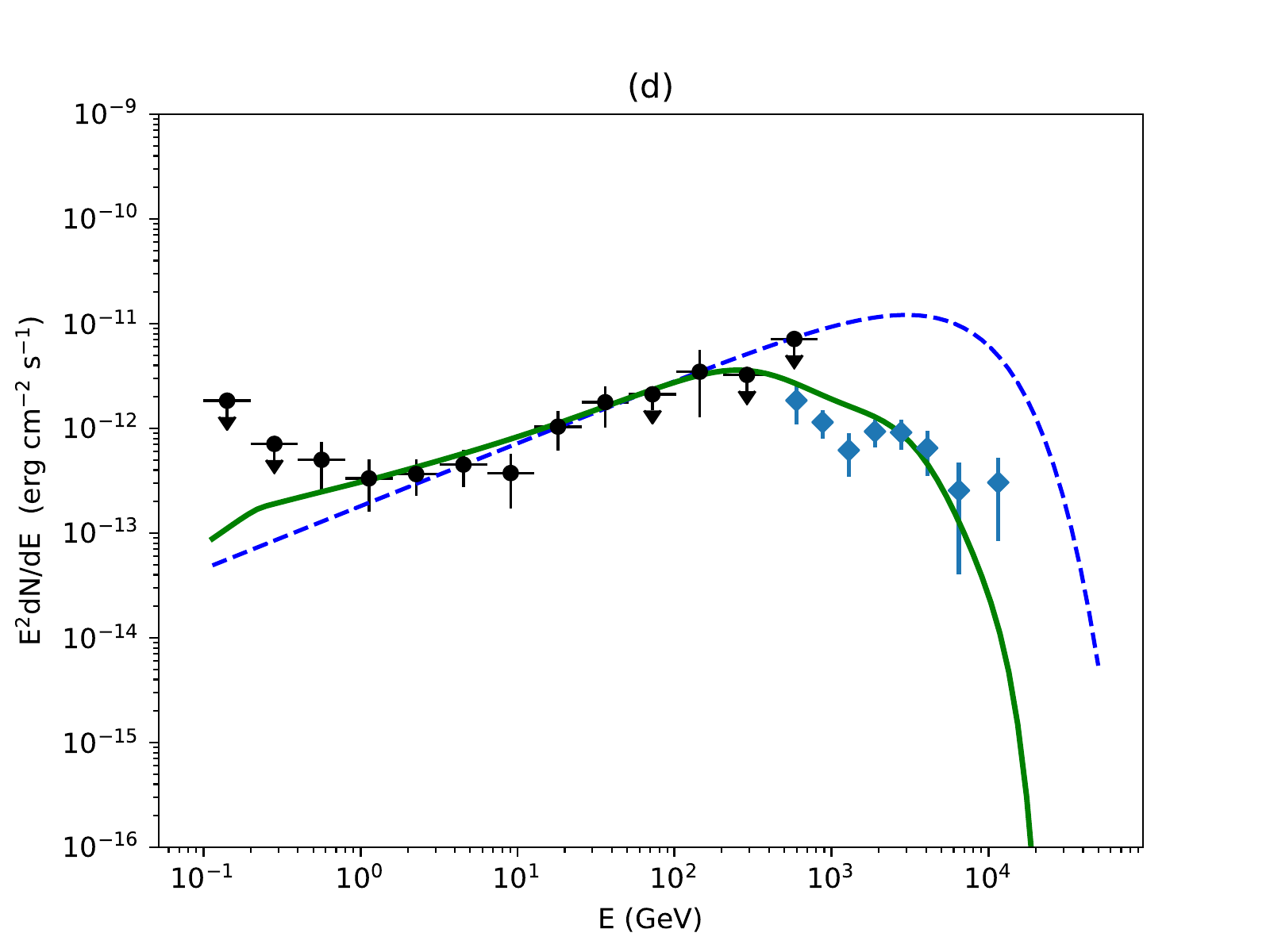}
\caption{From (a) to (d), same as Figures~\ref{c1}-\ref{c4} respectively, but with the EBL model of \cite{Finke10}.} 
\label{eblF}
\end{figure*}

In Figure~\ref{eblF}, we give the results with the EBL model of \cite{Finke10}. 
The other parameters are the same as that in Figures~\ref{c1}-\ref{c4}, respectively.
One can see that with this EBL model we cannot reproduce the HESS data below $\sim$ 1 TeV well, 
and the differences in the constraints on the IGMF caused by the two EBL models are negligible.

\end{document}